\documentclass[aps,pra,twocolumn,groupedaddress,showpacs]{revtex4}

\usepackage{graphicx}
\usepackage{graphicx}
\usepackage{amssymb}
\usepackage{amsmath}
\usepackage{color}
\newcommand{\be}{\begin{equation}}
\newcommand{\ee}{\end{equation}}
 
 \definecolor{BrickRed}{cmyk}{0,0.89,0.94,0.28}
\definecolor{MidnightBlue}{cmyk}{0.98,0.13,0,0.43}
\definecolor{DarkGreen}{rgb}{0,0.7,0.1}

\begin{document}

\title{On the Casimir effect between superconductors}


\author{Giuseppe Bimonte}

\affiliation{ Dipartimento di Fisica E. Pancini, Universit\`{a} di
Napoli Federico II, Complesso Universitario
di Monte S. Angelo,  Via Cintia, I-80126 Napoli, Italy}
\affiliation{INFN Sezione di Napoli, I-80126 Napoli, Italy}

\email{giuseppe.bimonte@na.infn.it}

\begin{abstract}

A recent experiment [R. A. Norte et al. Phys. Rev. Lett. {\bf 121}, 030405 (2018)] probed the variation of the Casimir force between two closely spaced  thin Al films, as they transition into a superconducting state, observing a null result. We present here  computations of the Casimir effect for superconductors, based on the Mattis-Bardeen formula for their optical response.  We  show that for the Al cavity used in the experiment the effect of the transition is over two hundred and fifty times smaller than the experimental sensitivity, in agreement with the observed null result. We demonstrate that a large enhancement of the effect can be achieved by  using  a cavity consisting of a Au  mirror and a superconducting NbTiN film. We estimate that the effect of the superconducting transition would be observable with the proposed Au-NbTiN configuration, if the sensitivity of the apparatus could be increased by an order of magnitude.

\end{abstract}


\maketitle

\section{Introduction}
\label{sec:intro}

One of the most spectacular manifestations of vacuum fluctuations of quantum fields is provided by the Casimir effect \cite{Casimir48}.  This is  the tiny force  acting between two discharged dielectric bodies,  which results from the modification of the spectrum of quantum and thermal fluctuations of the electromagnetic field in the region of space bounded by the two bodies. In his pioneering work, Casimir studied this phenomenon for the idealized case of two perfectly conducting plane-parallel mirrors at zero temperature. The investigation of the Casimir effect in real material media started with the  fundamental paper of Lifshitz \cite{lifs},  which presented a derivation of  the force between two plane-parallel dielectric slabs in vacuum, at finite temperature.   In recent years, intense experimental and theoretical efforts have been made to probe the dependence of the Casimir  force on the shapes and  material properties of the test bodies. For a review of the Casimir effect, and its perspective applications to nanotechnology the reader may consult several recent books and review articles \cite{book1,parse,book2,RMP,capasso,buh,woods,mehran}.  

Many experiments have now probed the Casimir  effect with test bodies made of  diverse materials, embedded in different media. Apart from two metallic  conductors in vacuum,  which still constitute the standard configuration,  experiments have been carried out with  semiconductors \cite{umar2006,umar2006bis,umar1}, conductive oxides \cite{iannuzzi,umar2,umar3}, magnetic materials \cite{bani2,ricardomag} and liquid crystals \cite{munday}. Experiments exist as well in which the bodies are  immersed  in  gases or in liquids \cite{liq,palas}. 

Another interesting class of candidate materials for Casimir experiments is represented by superconductors  \cite{ala1,ala2}.  The study of the Casimir effect in superconductors is indeed very interesting,  since these materials constitute an excellent arena \cite{bimontesuper} to  investigate  yet unresolved fundamental problems \cite{book2,RMP} about the influence of relaxation phenomena   on the strength of the Casimir force between metallic bodies.  Unfortunately, observing the influence of the superconducting transition on the Casimir effect is very difficult,  because on theory grounds one expects that the effect is extremely small. This can be understood  by considering that the transition modifies significantly  the optical properties of a superconductor only for frequencies of the order of $k_B T_c/\hbar$, where $T_c$ is the critical temperature. This region  represents only a very small fraction of the spectrum of frequencies that contribute to the Casimir interaction between two bodies at distance $a$.  The latter spectrum is known to stretch up to  the characteristic frequency  $\omega_c= c/a$, which for typical submicron separations is tens of thousands times larger than the frequency $k_B T_c/\hbar$ for classical BCS superconductors. In view of the difficulty of a direct force measurement,  in   \cite{ala1,ala2} we proposed an indirect approach, based on   observation of   the Casimir-induced shift of the critical magnetic field $H_{\rm c}$ of a thin superconducting film, constituting one of the two  plates of a rigid Casimir cavity. An experiment with an Al film based on this scheme,  placed an upper bound on the shift of the critical field not far from theoretical predictions \cite{superc,annalisa}. 

An alternative route to successful detection is represented by differential measurements, which offer the advantage of a far superior sensitivity   in comparison to absolute force measurements.  An experiment based on the observation of the differential Casimir force between a Au-coated sphere and the two sectors of a microfabricated plate, respectively made of superconducting Nb and Au,  was indeed proposed in \cite{bimonteiso}. The latter setup exploits the principle of isoelectronic differential measurements  \cite{bimoiso1,bimoiso2}, whose power in precision Casimir measurements has been demonstrated by a room temperature experiment \cite{ricardomag} with a microfabricated plate consisting of alternating Au and Ni sectors.  

More recently, an unpublished experiment  \cite{leiden}   measured the Casimir force between a Au-coated sphere with a radius $R=100\;\mu$m and a superconducting NbTiN film, with a critical temperature $T_c=13.6$ K.  The experimental data for room temperature showed  good agreement with theoretical predictions. The low-temperature data displayed however  an anomalous behavior,  due to an unexpected twenty percent increase in the measured  force, for which no explanation could be found. Apart from this, the experiment did not detect any change in the strength of the Casimir force across the superconducting transition,  and placed an upper bound of 2.6 \% on its magnitude. 

A promising on-chip platform for observing the Casimir force between superconductors has been described very recently in \cite{norte}. The apparatus consists of two  micro-fabricated Al-coated SiN parallel  strings,  having a length of 384 $\mu$m and a width of 926 nm. By application of a large tensile stress, the strings can be kept  perfectly parallel, at litographically determined fixed separations. Several cavities of different widths were realized on the same chip, the minimum separation being  of one hundred nm.  One of the two strings is attached to the movable mirror of an optomechanical cavity, whose resonance frequency is monitored by a laser.   The detection scheme  is based on the idea that when the system transitions to superconductivity, the resulting variation of the Casimir force between the Al strings should affect the mutual distance between the strings, thus determining  a change in the length of the the cavity  and therefore in its resonance frequency.  The experiment \cite{norte} provides a nice implementation the differential measurement scheme, since the  apparatus is sensitive to the variation $\Delta F(T)=F(T)-F(T_{\rm c})$ across the superconducting transition of the Casimir force $F(T) $ on the Al strings. Up to edge effects, the force $F(T) $  can be expressed as $F(T)=P(T) \times A $, where $P(T)$ is the unit-area Casimir force, i.e. the Casimir pressure, and $A$ is the area of the strings.  The above relation shows  that the measurement  of $\Delta F(T)$   is directly related  to the variation $\Delta P(T )= P(T )-P(T_{\rm c})$  of  the Casimir pressure across the critical temperature of the superconducting transition ($T_c=1.2$ K for Al).    The null result reported by the experiment sets  on the  magnitude of $\Delta P$ an upper bound of 6 mPa, which represents the sensitivity of the apparatus.     

In this paper we work out a detailed  theory of the Casimir effect in superconducting cavities. We  compute the variation $\Delta P(T)$ of the Casimir pressure for two distinct configurations of a superconducting planar cavity. In the first configuration, similarly to the experiment \cite{norte}, both plates are  made of the same superconductor, while in the second configuration, similarly to the experiment \cite{leiden}, one of the two superconducting plates is replaced by a Au mirror.  We model the frequency-dependent permittivity of the superconductor by the Mattis-Bardeen formula  \cite{mattis,tinkham}, which provides the best known description of the optical properties of superconductors. We present numerical results for  NbTiN and Al which are the superconductors used in the experiments  \cite{leiden}  and \cite{norte} respectively. It is important to note that optical measurements performed on NbTiN superconducting films \cite{hong} show excellent agreement with the local limit (so called dirty-limit) of the Mattis-Bardeen formula, providing strong support in favor of our theoretical model.   Our computations show that for the Al  cavity  used in the experiment \cite{norte}  the magnitude of $\Delta P$ is over two hundred and fifty times smaller than the experimental sensitivity. Our results, while in agreement with the null result reported by the experiment, make it  unlikely that the effect of the superconducting transition can be observed with an Al cavity.  We find however that the magnitude of  $\Delta P(T)$ can be  enhanced by a factor of fifteen, by  considering a cavity  composed by a Au mirror  and a NbTiN film, having a thickness larger than two hundred nm.  The enhancement factor increases to thirty-four if the separation $a$ is decreased from 100 nm to 60 nm. This is an encouraging result, since it shows that the effect would be detectable with a Au-NbTiN cavity, if the sensitivity of the apparatus  could be improved by one order of magnitude.

The plan of the paper is as follows: in Sec. II we review the general formalism for computing the Casimir pressure between two superconducting plates, and we present the models we use to describe their optical properties. In Sec. III we present the results of our numerical computations. In Sec. IV we present our conclusions. Finally, in the Appendix we  provide the explicit formula for the analytic continuation to the imaginary frequency axis of the Mattis-Bardeen formula for the frequency dependent conductivity of  BCS
superconductors.

\section{General formalism for the Casimir pressure}

We consider a  Casimir cavity, formed by two plane-parallel homogeneous and isotropic dielectric plates at temperature $T$, separated by an empty gap of width $a$.  We denote by $\epsilon^{(k)}(\omega)$, $k=1,2$ their respective (complex) permittivities (we only consider non magnetic materials, and thus we set $\mu^{(1)}= \mu^{(2)} \equiv 1$).
According to Lifshitz formula \cite{lifs}, the Casimir pressure $P(a,T)$ among the plates can be expressed as (negative pressures correspond to attraction):
$$
P(a,T)=-\frac{k_B T}{\pi}\sum_{l=0}^{\infty}\;\!\!' \int_0^{\infty}dk_{\perp} k_{\perp} q_l 
$$
\be
\times \sum_{\alpha} \left[\frac{e^{2 a q_l}}{r^{(1)}_{\alpha}(i \xi_l, k_{\perp}) r^{(2)}_{\alpha}(i \xi_l, k_{\perp})}-1 \right]^{-1}\;,\label{lifs}
\ee  
where $k_B$ is Boltzmann constant, $k_{\perp}$ is the in-plane momentum, the prime in the sum indicates that the  $l=0$ term is taken with weight one-half, $\xi_l= 2 \pi l k_B T/\hbar$ are the imaginary Matsubara frequencies, $q_l=\sqrt{\xi_l^2/c^2+k_{\perp}^2}$, and the sum over $\alpha={\rm TE, TM}$ is taken over the independent states of polarization of the electromagnetic field, i.e. transverse magnetic ($\rm TM$) and transverse electric  ($\rm TE$). Finally, the symbols $r^{(k)}_{\alpha}(i \xi_l, k_{\perp}) $ denote the Fresnel reflection coefficients of the $k$-th slab:
\be
r^{(k)}_{\rm TE}(i \xi_l, k_{\perp}) =\frac{q_l-s^{(k)}_l}{q_l+s^{(k)}_l}\;,\label{TE}
\ee
\be
r^{(k)}_{\rm TM}(i \xi_l, k_{\perp}) =\frac{\epsilon^{(k)}_l q_l-s^{(k)}_l}{\epsilon^{(k)}_l\,q_l+s^{(k)}_l}\;,\label{TM}
\ee
where $s^{(k)}_l=\sqrt{\epsilon^{(k)}_l \xi_l^2/c^2+k_{\perp}^2}$, and $\epsilon^{(k)}_l \equiv \epsilon^{(k)}(i \xi_l)$. If instead of thick homogeneous slabs, one considers more complex mirrors constituted by  plane-parallel metallic  films deposited on some substrate,  the corresponding Casimir pressure can still be computed by the general Lifshitz formula Eq. (\ref{lifs}), provided that the Fresnel reflection coefficients Eqs. (\ref{TE}-\ref{TM})  are replaced by the reflection coefficients of the layered mirrors \cite{book2}.  We shall consider two distinct confìgurations for our system: in the first one, both plates are made of the same superconducting material. Concretely, we shall consider two superconductors, i.e. Al (which is the superconductor used in the experiment \cite{norte}), and NbTiN (which is the superconductor used in the experiment \cite{leiden}). The corresponding configurations shall be denoted as Al-Al and NbTiN-NbTiN, respectively.  The respective Casimir pressures  are obtained by substituting into Lifshitz formula the permittivities of Al or NbTiN, respectively: $\epsilon^{(1)}_l= \epsilon^{(2)}_l=\epsilon^{(\rm Al/ NbTiN)}(i \xi_l)$. In the second configuration, one of the two superconducting plates  is replaced by a   Au mirror. This second configuration shall be analyzed in detail only for the case of NbTiN,  and we shall denote it as  the Au-NbTiN configuration.  The corresponding Casimir pressure is obtained by setting into Eq. (\ref{lifs})   $\epsilon^{(1)}_l=  \epsilon^{(\rm Au)}(i \xi_l)$ and  $\epsilon^{(2)}_l=   \epsilon^{(\rm NbTiN)}(i \xi_l)$. 

To compute the Casimir pressure, one needs the  permittivities $\epsilon^{(k)}_l$ of the materials constituting the plates. In a concrete experimental situation, one would  ideally like to measure the optical data of the used samples, for the experimental values of the temperature.   The permittivities  $\epsilon^{(k)}_l$ for the physically  inaccessible imaginary frequencies $i \xi_l$ would then be computed on the basis of the optical data, using Kramers-Kronig dispersion relations \cite{book2}. In order to obtain a precise theoretical estimate of the Casimir pressure $P(a,T)$ for a separation $a$, it is in principle necessary to  know the optical data for all frequencies lower than ten or twenty times the characteristic cavity frequency $\omega_c=c /2 a$ \cite{book2}. For $a=100$ nm, $\omega_c=1.5 \times 10^{15}$ rad/s.  

It is fortunate that in the problem at hand  we do not really need this much information about the optical properties of the materials. Indeed,  the quantity that interests us is not  the Casimir pressure $P(a,T)$ at a single temperature, but rather its {\it variation} $\Delta P(a;T)$ across the critical temperature $T_{\rm c}$:
\be
\Delta P(a;T)=P(a,T )-P(a,T_{\rm c})\;,
\ee      
where $T < T_{\rm c}$.   Now, it is known \cite{tinkham} that  the superconductive transition  affects significantly the optical properties of a superconductor only for frequencies corresponding to  photon energies smaller than (a few times)  the BCS gap  $\Delta(0)$. From BCS theory \cite{tinkham} one knows that $\Delta(0) = 1.76 k_B T_{\rm c}$. For Al ($T_{\rm c}=1.2$ K) this gives $\Delta(0)=1.8 \times 10^{-4}$ eV, while for NbTiN ($T_{\rm c}=13.6$ K)  $\Delta(0)=2.1 \times 10^{-3}$ eV.     
For these small photon energies the optical response of a normal metal is dominated by intraband transitions. The latter can be phenomenologically described by a Drude-model   dielectric function of the form
\be
\epsilon(i \xi)=\epsilon_{0}+ \frac{\Omega^2}{\xi (\xi + \gamma)}\;,\label{drude}
\ee
where the contribution from core interband transitions has been included in $\epsilon_0$.   
Here  $\Omega$ is the  plasma frequency for intraband transitions, and $\gamma$ is the relaxation frequency.   To compute $\Delta P$ we have used the simple Drude model in Eq. (\ref{drude}) to describe the permittivity of Au, as well as the permittivity of the superconductors in the normal state. In our computations we have  neglected the temperature dependence of both the plasma frequency $\Omega$ and of the core-electron permittivity $\epsilon_0$, and thus we used their room temperature values.  The relaxation frequency is instead temperature dependent, and in general it decreases as the temperature is decreased. At cryogenic temperatures $\gamma$ approaches a constant sample dependent residual value. Following the standard convention, we express the  residual relaxation frequency in terms of  the corresponding  room temperature frequency $\gamma_0$ by the formula $\gamma=\gamma_0/RRR$ where RRR is the residual resistance ratio.  The values of the parameters were chosen as follows. For Au, we used the standard values $\Omega =9$ eV/$\hbar$ and  $\gamma_0= 35$ meV/$\hbar$ \cite{book2}, while from the tabulated optical data \cite{Palik}  we obtained $\epsilon_0=6.3$. For Al, we used  $\Omega=13$ eV/$\hbar$,  $\gamma_0= 100$ meV/$\hbar$,  and $\epsilon_0=1.03$  \cite{Palik}. Finally, for NbTiN we used the values quoted in \cite{leiden} i.e.  $\Omega=5.33$ eV/$\hbar$ and  $\gamma_0= 0.465$ eV/$\hbar$. In \cite{leiden} the optical data of the used NbTiN films were determined by ellipsometry in the frequency range from 1.89$\times 10^{11}$ rad/s to 1.13$\times 10^{16}$ rad/s, both at room temperature and at 16 K. The optical data were afterwards fitted by a Lorentz-Drude model with four oscillator terms.  Unfortunately the values of the corresponding  parameters  were not reported explicitly. We are thus unable to provide a value for the contribution of core electrons for this material. We have verified however that the pressure variation $\Delta P(a;T)$ remains practically unchanged when the value $\epsilon_0$ for NbTiN is varied in the interval from  one to ten. The value of the RRR parameter depends on the sample preparation procedure, and therefore it cannot be fixed a priori.  For the NbTiN sample used in the experiment \cite{leiden},  the fit to the optical data at 16 K  gave $\gamma=0.415$, which corresponds  to RRR=1.12.  To probe the sensitivity of the pressure  variation $\Delta P$ on this parameter,  in our computations we varied its value  in the interval from one to ten.  
\begin{figure}
\includegraphics [width=.9\columnwidth]{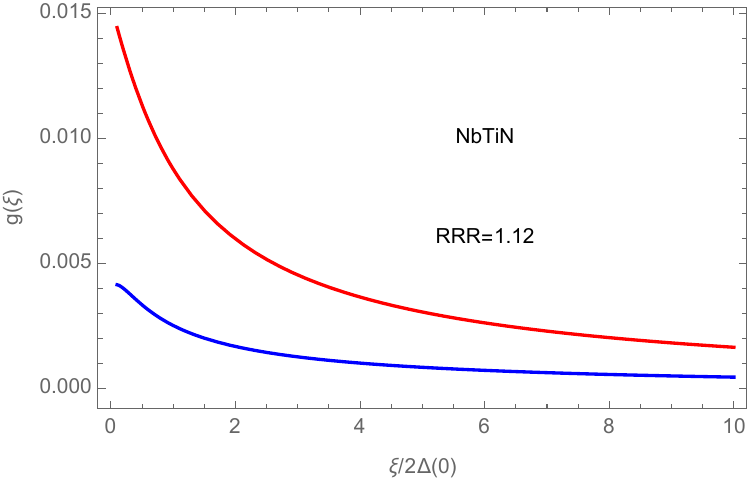}
\caption{\label{Fig1bis}  Plot of  $g(\xi)$ (see Eq. (\ref{MB})) for NbTiN (RRR=1.12), versus $\xi/2 \Delta(0)$  for $T/T_{\rm c}=0.9$ (blue line), and $T/T_{\rm c}=0.1$ (red line).}
\end{figure}

\begin{figure}
\includegraphics [width=.9\columnwidth]{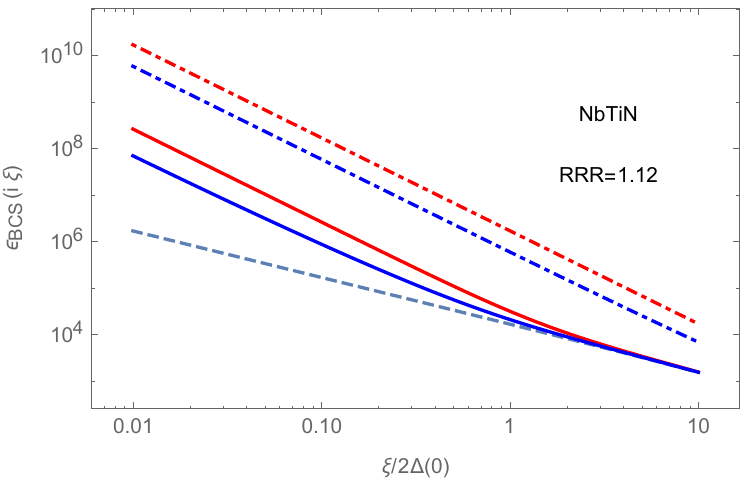}
\caption{\label{Fig1}  BCS permittivity of NbTiN (RRR=1.12)  versus $\xi/2 \Delta(0)$  for $T/T_{\rm c}=0.9$ (solid blue line) and $T/T_{\rm c}=0.1$ (solid red line). The dashed line  shows the Drude permittivity Eq. (\ref{drude}). The dot-dashed lines show the permittivity for the Casimir-Gorter two-fluid model, for $T/T_{\rm c}=0.9$ (dot-dashed blue line), and $T/T_{\rm c}=0.1$ (dot-dashed red line)}
\end{figure}

\begin{figure}
\includegraphics [width=.9\columnwidth]{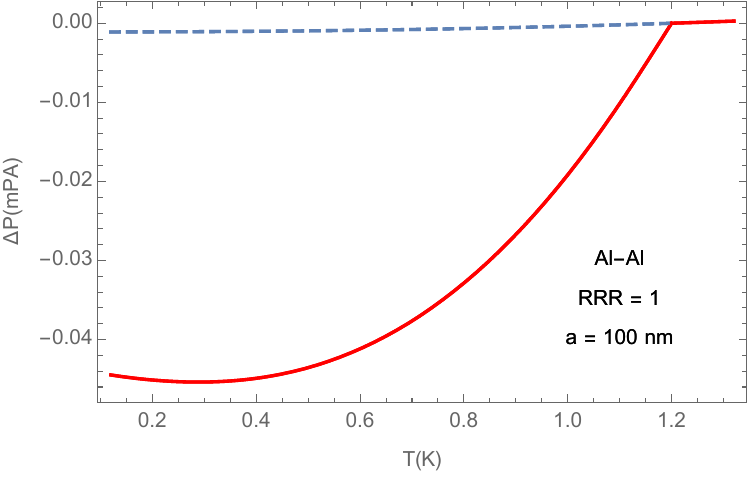}
\caption{\label{Fig3}   Variation of the Casimir pressure across the superconducting transition   of  an Al cavity with thick walls, versus temperature (in K).  The dashed line represents the variation of the Casimir pressure in the absence of the transition.}
\end{figure}

Next, we describe the model for the permittivity of the superconductors. For this we rely on the Mattis-Bardeen formula \cite{mattis} for the conductivity $\sigma$, which is known to provide an accurate representation of the optical response of  BCS superconductors \cite{tinkham}. In its general form, the Mattis Bardeen formula depends both on the frequency $\omega$ and the wavevector $q$,  since superconductors display spatial dispersion.  However, the $q$-dependence is negligible in  the so-called dirty limit $\ell / \xi_0 \ll 1$, where $\ell=v_F/\gamma$ is the mean free path, and $\xi_0=\hbar v_F/\pi \Delta(0)$ is the correlation length, with $v_F$ the Fermi velocity. The dirty limit condition is well satisfied by both  Al ($\ell / \xi_0=5.7 \times 10^{-3}$) and NbTiN  ($\ell / \xi_0=1.4 \times 10^{-2}$).  This is confirmed by optical measurements of NbTiN films in the THz region, that are in excellent agreement with the local dirty-limit of the Mattis-Bardeen formula \cite{hong}. The analytic continuation of the Mattis-Bardeen formula to the imaginary frequency axis has been worked out in \cite{bimonteBCS}, where it is shown that $\sigma_{\rm BCS}(i \xi)$ can be conveniently decomposed as:
\be
\sigma_{\rm BCS}(i \xi)=\frac{\Omega^2}{4 \pi}  \left[\frac{1}{(\xi+\gamma)}+  \frac{g(\xi;T)}{\xi}\right]\;,\label{MB}
\ee
The first term between the square brackets on the r.h.s. of the above Equation coincides with the familiar Drude contribution to the conductivity of a normal metal, while the second term represents the BCS correction. The explicit expression of the function $g(\xi)$  is given in the Appendix.  Here is a brief summary of its main properties.  The function $g(\xi)$ is different from zero only for $T< T_{\rm c}$, and vanishes identically for $T \rightarrow T_{\rm c}$. For $T< T_{\rm c}$, it is a positive and monotonically decreasing function of $\xi>0$, approaching a finite value $g(0)<1$  for $\xi \rightarrow 0$, and going to zero for $\xi\rightarrow \infty$. Its value depends parametrically on the temperature-dependent BCS gap $\Delta$ as well as on the relaxation frequency $\gamma$. In addition to that, $g(\xi)$ has an explicit dependence on the temperature. For   small $\xi$  the function $g(\xi)$ has the expansion:
\be
g(\xi;T)= \omega_s^2(T)+B(T) \xi \log(\Delta/\hbar\xi)+ o(\xi)\,,
\ee
where $\omega_s(T)$ represents the (normalized) effective superfluid plasma frequency.  A plot of the function $g(\xi;T)$ for NbTiN (RRR=1.12) is shown in Fig. \ref{Fig1bis} for  $T/T_{\rm c}=0.9$ (blue line) and for $T/T_{\rm c}=0.1$ (red line).
By adding the contribution of core electrons, we thus  arrive at the following formula for the permittivity of the superconductor:
\be
\epsilon_{\rm BCS}(i \xi)=\epsilon_{0}+4 \pi\frac{\sigma_{\rm BCS}(i \xi)}{\xi}=\epsilon_{0}+ \frac{\Omega^2}{\xi }\left[\frac{1}{\xi+ \gamma}+   \frac{g(\xi;T)}{\xi}\right]\;.\label{BCS}
\ee
The BCS  term proportional to $g(\xi;T)$ in the expression of $\epsilon_{\rm BCS}$  can be interpreted as a plasma-model contribution, with an effective $\xi$-dependent plasma frequency $\Omega_{\rm eff}(\xi)=\Omega \sqrt{g(\xi;T)}$.
In Fig. \ref{Fig1} we show  logarithmic  plots of the BCS permittivity  of NbTiN  as a function of $\xi/2 \Delta(T)$, for $T/T_{\rm c}=0.9$ (blue line) and for $T/T_{\rm c}=0.1$ (red line).  The dashed line corresponds to the  Drude permittivity Eq. (\ref{drude}). The figure shows that the BCS permittivity  approaches the Drude permittivity for $\xi/2 \Delta(0) \gg 1$.

It is interesting to compare the BCS formula for the permittivity with the old-fashioned Casimir-Gorter two fluid-model \cite{gorter,tinkham}. According to this model a fraction $n_{\rm s}(T)$ of the conduction electrons contributes to the supercurrent, while the remaining fraction $n_{\rm n}(T)=1-n_{\rm s}(T)$ remains normal. Superconducting electrons  behave as a dissipationless plasma, while  normal
electrons are described by the usual dissipative Drude model. Core electron remain unaltered. According to this simple physical picture, the  permittivity of the two-fluid model  is written as:
\be
\epsilon(i \xi)=\epsilon_0+(1-n_{\rm s}(T))\frac{\Omega^2}{\xi(\xi+\gamma)}+ n_{\rm s}(T)\frac{\Omega^2}{\xi^2}\;.\label{CG}
\ee
The fraction $n_{\rm s}(T)$ of superconducting electrons follows the Casimir-Gorter law:
\be
n_{\rm s}(T)=\left[1-\left(\frac{T}{T_{\rm c}} \right)^4\right] \Theta (T_{\rm c}-T)\;,
\ee
where $\Theta(x)$ is the Heaviside step-function: $\Theta(x)=1$ for $x >0$, and  $\Theta(x)=0$ for $x \le 0$. In Fig. \ref{Fig1} we show plots of the two-fluid model for NbTiN, for $T/T_{\rm c}=0.9$ (blue dot-dashed line) and for $T/T_{\rm c}=0.1$ (red dot-dashed line). Comparison with the BCS permittivity (solid lines) shows that the two-fluid model  overestimates   the permittivity of a superconductor by a very large factor. We note that  the two fluid model was used in \cite{leiden} to compute the Casimir force between superconductors.

\section{Numerical computation of $\Delta P$}

In this Section we present our numerical computations of the pressure variation $\Delta P(a;T)$, based on the expressions of the permittivity described in the previous Section. 

We consider first a Casimir cavity constituted by  two thick plates made of Al, which is the  superconductor used in the experiment \cite{norte}. In Fig. \ref{Fig3}  the corresponding  $\Delta P(a;T)$ is plotted versus the temperature $T$ (in K),  for the separation $a=100$ nm   which was the minimum separation probed in the experiment. We took RRR=1. For comparison, we show in the same Figure the variation of the pressure that would obtain in the absence of the transition (dashed line). We see that  the solid curve lies below the dashed one, in accordance with one's expectation that the superconducting transition determines an increase in the Casimir attraction with respect  to the normal state, since superconductors are better reflectors than normal metals. The magnitude of $\Delta P$ is seen to be smaller than 0.05 mPa at all temperatures below $T_{\rm c}$.  To get a feeling of how small an effect this represents, we note that   the magnitude of the Casimir force $P(T_{\rm c})$ at the critical temperature is estimated to be of 6.8 Pa (this value was computed using the simple representation Eq. (\ref{drude}) for the permittivity of Al, and must be just considered as an approximate estimate. A more accurate estimate would require a better description of core electrons).   Using this estimate, we obtain  $\Delta P/|P(T_{\rm c})|<$7$\times 10^{-6}$ 
across the transition.

The Casimir cavity  used the experiment \cite{norte} consisted of two identical layered plates, each consisting of an Al film  with a thickness $w=18$ nm, deposited on a SiN substrate. 
\begin{figure}
\includegraphics [width=.9\columnwidth]{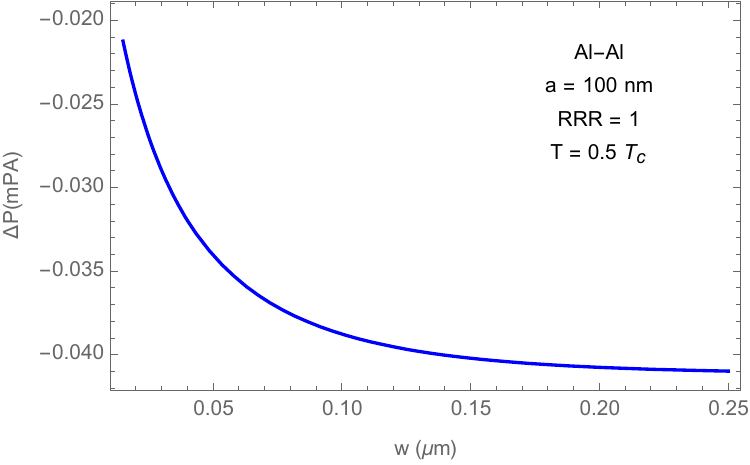}
\caption{\label{Fig4} Variation of the Casimir pressure across the superconducting transition  of an Al cavity consisting of two identical Al films of thickness $w$ deposited on a SiN substrate,  versus films thickness $w$ (in $\mu$m).}
\end{figure}
To determine  how the thickness $w$ of the Al films influences $\Delta P$, it is necessary to replace in Lifshitz formula the Fresnel reflection coefficients for a thick Al slab Eqs. (\ref{TE}-\ref{TM}) by those for the layered Al-SiN plate \cite{book2}.   The resuls of this computation are shown in Fig. \ref{Fig4}.  We see that the thick-plate limiting value of 0.041 mPa is  nearly reached for a film thickness of 250 nm,  but for a thickness of 18 nm the  magnitude of $\Delta P$ decreases to  0.023 mPa.  Recalling that the experiment \cite{norte} has an estimated sensitivity of 6 mPa, we see that  the theoretical pressure variation for the Al cavity is over  two hundred and fifty times smaller than the sensitivity.  While this is consistent with the null result reported by the experiment, it makes one think that observation of the effect  with the Al cavity is hardly possible in the near future. 
 
\begin{figure}
\includegraphics [width=.9\columnwidth]{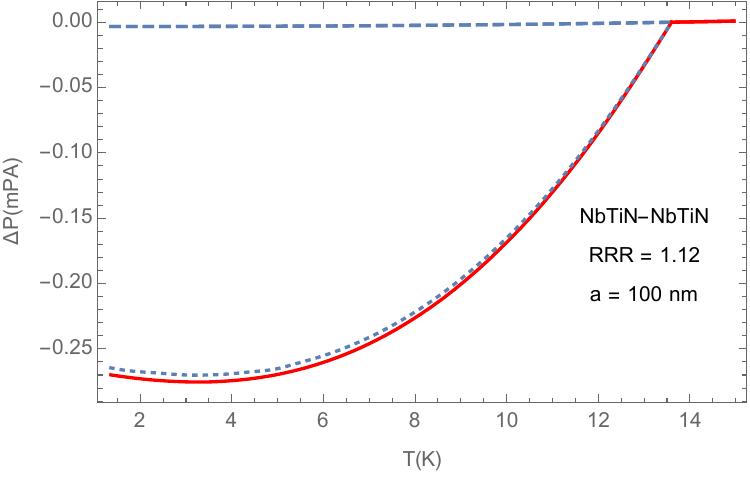}
\caption{\label{Fig5}   Variation of the Casimir pressure across the superconducting transition  of a NbTiN cavity, versus temperature (in K).  The dashed line represents the variation of the Casimir pressure in the absence of the transition. The solid and dotted lines correspond, respectively, to taking $\epsilon_0=1$ and $\epsilon_0=10$ in the permittivity of NbTiN (see Eq. (\ref{drude})).}
\end{figure}

Our computations predict that  a significant increase in the magnitude of the pressure variation can be achieved by using NbTiN in the place of Al. This is demonstrated by Fig. \ref{Fig5}, which displays the pressure variation for two thick NbTiN plates at a separation $a=100$ nm, versus the temperature $T$ (RRR=1.12). As we said earlier, we could not find in the literature enough information on the optical  properties of NbTiN, to fix the value of  $\epsilon_0$ in Eq. (\ref{BCS}).  For this reason, we repeated the computations using two widely different values for $\epsilon_0$.  It is fortunate that the pressure variation is insensitive to the contribution of core electrons, as it can be seen  from Fig. \ref{Fig5} where the solid and dotted  lines  correspond  to $\epsilon_0=1$ and $\epsilon_0=10$, respectively. The weak dependence of $\Delta P$ on $\epsilon_0$ is explained by the fact that 
the pressure variation is determined by the optical response of the materials at frequencies of the order the thermal frequency $k_B T_{\rm c}/\hbar$, for which the Drude term is overwhelmingly  large compared to $\epsilon_0$.  Comparison of Fig \ref{Fig5} with Fig. \ref{Fig3} shows that the variation of the Casimir pressure for a NbTiN cavity  is five times larger than the corresponding  variation for an Al cavity. 

\begin{figure}
\includegraphics [width=.9\columnwidth]{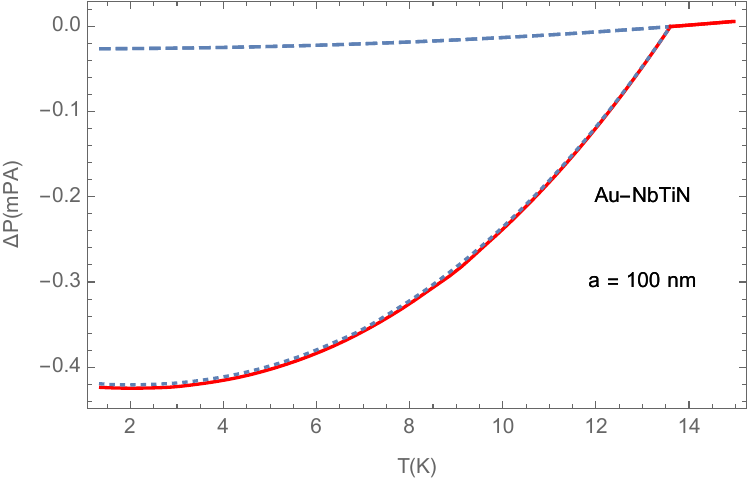}
\caption{\label{Fig6}   Variation of the Casimir pressure  across the superconducting transition of a Au-NbTiN cavity, versus temperature (in K).  The dashed line represents the variation of the Casimir pressure in the absence of the transition. The solid and dotted lines correspond, respectively, to taking $\epsilon_0=1$ and $\epsilon_0=10$ in the permittivity of NbTiN (see Eq. (\ref{drude})). Data are for RRR$_{\rm NbTiN}=1.12$ and RRR$_{\rm Au}=1$.} 
\end{figure}

\begin{figure}
\includegraphics [width=.9\columnwidth]{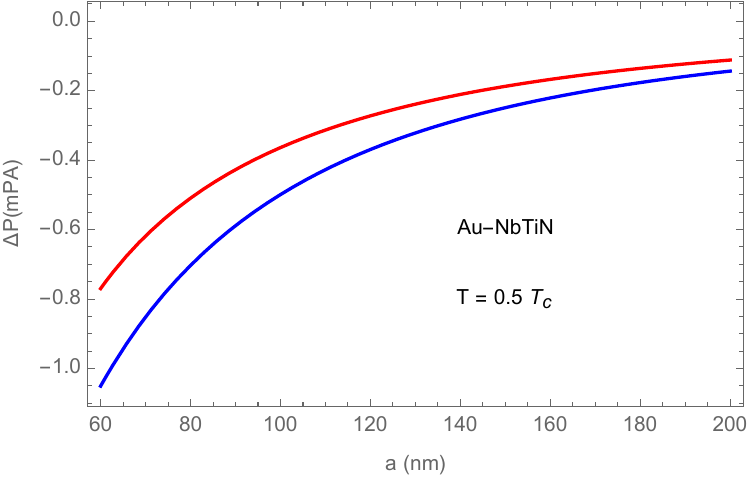}
\caption{\label{Fig7}   Variation of the Casimir pressure  across the superconducting transition of a Au-NbTiN cavity, versus separation $a$ (in nm).   The red and blue lines correspond, respectively, to residual resistance ratios RRR=1.12 and RRR=5 for the NbTiN film. In both cases RRR$_{\rm Au}$=1.}
\end{figure}

\begin{figure}
\includegraphics [width=.9\columnwidth]{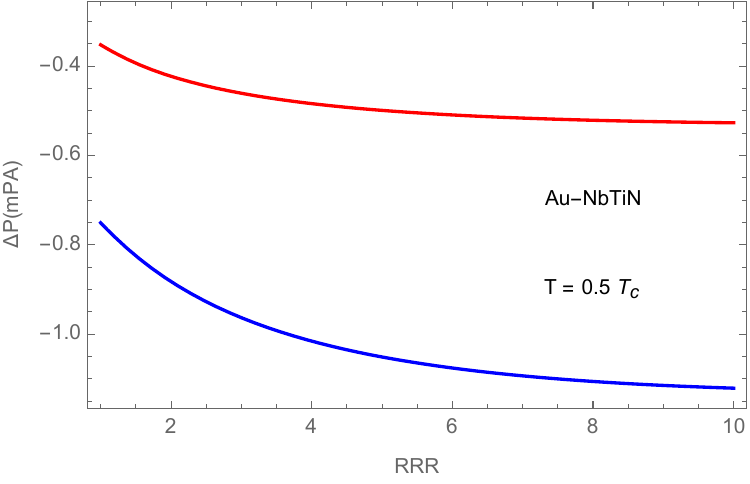}
\caption{\label{Fig8}   Variation of the Casimir pressure across the superconducting transition  of a Au-NbTiN cavity, versus the RRR of the  NbTiN plate. The RRR of the Au plate is fixed to one. The red and blue lines correspond respectively to the separations $a=100$ nm and $a=60$ nm .}
\end{figure}

\begin{figure}
\includegraphics [width=.9\columnwidth]{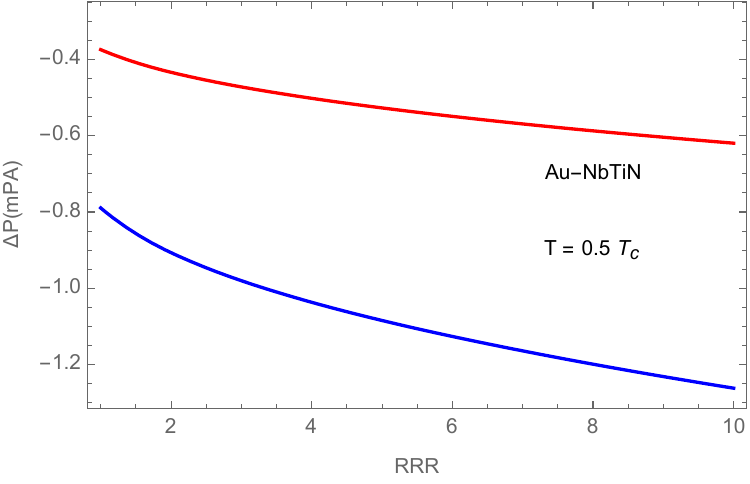}
\caption{\label{Fig9}   Variation  of the Casimir pressure across the superconducting transition of a Au-NbTiN cavity, versus the RRR of the Au plate. The RRR of the NbTiN plate is fixed to RRR=1.12. The red and blue lines correspond respectively to the separations $a=100$ nm and $a=60$ nm .}
\end{figure}

\begin{figure}
\includegraphics [width=.9\columnwidth]{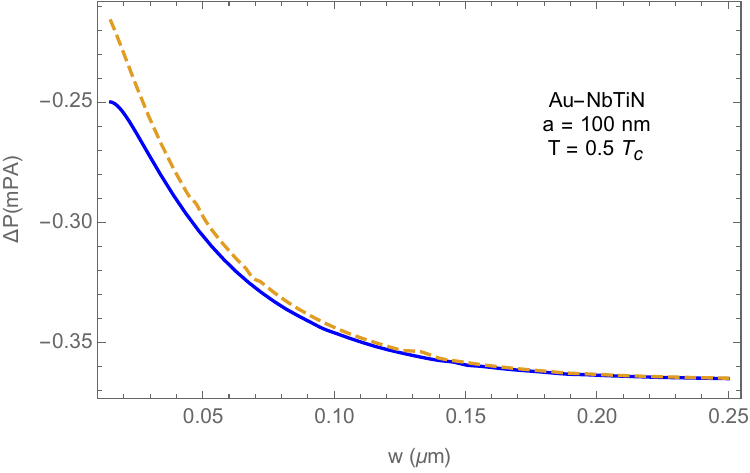}
\caption{\label{Fig10}   Variation  of the Casimir pressure  across the superconducting transition of a cavity formed by a thick Au mirror and a NbTiN film of thickness $w$,  versus film thickness $w$ in $\mu$m. Shown data are for RRR$_{\rm NbTiN}=1.12$ and RRR$_{\rm Au}=1$. The solid  and dashed lines correspond, respectively, to a free standing NbTiN film, and to a film deposited on a substrate having a static permittivity equal to ten.}
\end{figure}

A further increase of the pressure variation can be achieved by replacing one of the NbTiN plates by a Au  mirror. This was the combination of materials adopted in the unpublished  experiment \cite{leiden}. We assume in what follows that the thickness of the Au coating of the first mirror is larger than 200 nm. This ensures that  for the purposes of the Casimir effect that mirror can be considered as equivalent to an infinitely thick Au slab \cite{book2}. We note that Ref. \cite{leiden} does not provide data for the RRR of Au at 16 K. In our computations we take $RRR_{\rm Au}=1$.  It can be seen from Fig. \ref{Fig6}  that  for $a=100$ nm  the maximum   variation pressure for the Au-NbTiN cavity  has  a magnitude of 0.42 mPa, which is nine times larger than the corresponding maximum pressure variation of the Al cavity (see Fig. \ref{Fig3}). In Fig. \ref{Fig7} we show the  pressure variation  of the Au-NbTiN cavity as a function of the separation $a$ (in nm), for $T/T_{\rm c}=0.5$. 
The red and blue curves  correspond to RRR$_{\rm NbTiN}$=1.12 and RRR$_{\rm NbTiN}$=5, respectively. In Fig. \ref{Fig8} the pressure variation is displayed versus the residual resistance ratio RRR of the NbTiN film, for the two separations $a=100$ nm (red curve) and $a=60$ nm (blue curve).  In Fig. \ref{Fig9} the pressure variation is displayed versus the residual resistance ratio RRR of the Au film, for the two separations $a=100$ nm (red curve) and $a=60$ nm (blue curve). For both curves, the RRR of the NbTiN film has the fixed value RRR$_{\rm NbTiN}$=1.12. We note that by increasing the value of  RRR for the Au plate,  it is possible to obtain a significant increase in   the magnitude of  $\Delta P$. This indicates that it would be  beneficial to realize a Au mirror having a long mean free path for the electrons. 
Finally, in Fig. \ref{Fig10} we display the pressure variation of the cavity formed by a thick Au mirror and a NbTiN film of thickness $w$,  as a function of the film thickness $w$ (in $\mu$m), for a fixed separation $a=100$ nm. The pressure variation is moderately dependent on the properties of the substrate of the superconducting film. We verified this by comparing the results for a free-standing film (solid line of Fig. \ref{Fig10})  with those for a substrate having a static permittivity equal to ten (dashed line in Fig. \ref{Fig10}). The influence of the substrate of course decreases for thicker films.  The plot shows that NbTiN films with a thickness larger than two hundred nm are essentially undistinguishable from an infinitely thick slab.

The important conclusion that can be drawn from the computations described above,  is that a large enhancement of the pressure variation $\Delta P$  can be achieved by replacing the thin Al plates   used in the experiment \cite{norte}, by  a cavity composed by a thick Au mirror and a NbTiN film having a thickness larger than two hundred nm. To get a quantitative idea of the magnitude of the  enhancement that can be achieved in this way, consider as an example a cavity with a width $a=100$ nm, at a temperature $T=0.5\, T_{\rm c}$.  For the Al cavity of \cite{norte}, one gets  $\Delta P=-0.023$ mPa  while for the Au-NbTiN cavity (with RRR$_{\rm Au}$=1 and RRR$_{\rm NbTiN}$=1.12) one finds $\Delta P=-0.36$ mPa.  While this figure represents a  15.8-fold enhancement with respect to the Al cavity, it is still 16.5 times smaller than the sensitivity of 6 mPa. One can get closer to the sensitivity threshold by decreasing the separation $a$. For example, going down to a=60 nm, one gets $\Delta P=-0.77$ mPa, which is 7.8 times smaller than the sensitivity. The remaining gap can be partly filled by improving the mean free path $\ell$ of the Au mirror. If Au mirrors with RRR=3 can be made, that would give $\Delta P=-0.98$ which is just 6.1 times smaller than the sensitivity. This shows that the effect of the transition would be observable with the Au-NbTiN cavity if the sensitivity of the apparatus could be improved by only one order of magnitude.

As a final remark, we note that  in the unpublished experiment using a superconducting Au-NbTiN cavity  \cite{leiden}, the Casimir pressure  was computed  using the Casimir-Gorter two-fluid model Eq.(\ref{CG}) for the permittivity of the superconductor. Unfortunately, the results obtained in this way are not quite correct.  Using this model, the authors estimated that  for $a=100$ nm, the variation of the Casimir pressure $\Delta P(a;T)$  for $T \ll T_{\rm c}$ was of -265 mPa, corresponding to a 5.1 \% fractional change $\Delta P(T)/|P(T_{\rm c})|$ of the Casimir pressure.  The corresponding  variation $\Delta P(a;T)$  obtained by us using the BCS permittivity  is of -0.42 mPa (see Fig. \ref{Fig6}),  which amounts to a pression fractional change $\Delta P(T)/|P(T_{\rm c})|=8 \times 10^{-5}$. We thus see that the two-fluid model overestimates the magnitude of the pressure variation by a factor larger than 600.    We note also that the prediction of a  5.1 \% change in the pressure is   in disagreement with the experimental bound of 2.5 \%, while of course the prediction of the BCS model  is consistent with it.     

\section{Conclusions}

A much debated problem in the theory of the Casimir effect is the role of relaxation phenomena of free charge carriers in Lifshitz theory \cite{book2,RMP}. Different prescriptions have been proposed in the  literature to compute the Casimir  force between conducting test bodies, that go by the names of Drude and plasma prescriptions \cite{book2,RMP}. Superconductors offer a unique possibility to investigate this problem \cite{bimontesuper}. Unfortunately, it is very difficult to   probe the influence of  the superconducting transition  on the Casimir force, because the effect of the transition is expected to be very small \cite{bimontesuper}. A recent experiment with thin superconducting Al films reported a null result \cite{norte}. 

In this paper we have developed a detailed theory for the Casimir effect with superconducting plates.   Our analysis  relies on the Mattis-Bardeen formula for the frequency-dependent conductivity of BCS superconductors, which  represents the best known theoretical description of the optical properties of superconductors.  We  performed numerical computations for Al and for NbTiN, which are the superconductors used in the experiments \cite{norte} and \cite{leiden}, respectively. The excellent agreement with the Mattis-Bardeen formula     demonstrated by recent optical measurements on superconducting NbTiN \cite{hong}, lends strong support to the  validity of our theoretical analysis.   We estimate that for the Al cavity used in the experiment \cite{norte},  the magnitude of the variation of the Casimir pressure across the transition is over two hundred and fifty times smaller than the sensitivity of the experiment. This result, while consistent with the observed null result, makes it unlikely that the effect of the superconducting transition can be observed with an Al cavity.  We find however that the expected signal can be enhanced by a factor of fifteen by substituting the thin Al films used in \cite{norte} with a Casimir cavity constituted by a Au mirror  and a NbTiN superconducting film, having a thickness larger than two hundred nm. The   enhancement factor  increases to thirtyfour times, if the  width of the cavity is decreased  from  100 nm to 60 nm.  According to our computations, a further improvement is possible by using a Au mirror with a long mean free path for the electrons.   Our analysis shows that  the  effect of the transition to superconductivity  would be observable with the Au-NbTiN cavity,  if the sensitivity of the apparatus used in \cite{norte} could be increased by  one order of magnitude.
  
\acknowledgments

The author thanks  R.  A. Norte for  useful discussions on the experiment \cite{norte}.  

 .
\appendix*

\section{Expression of the function $g(\xi)$}

In this Appendix we display the explicit expression of the function $g(\xi)$ that enters in Eq. (\ref{MB}), providing  the analytic continuation  to the imaginary frequency axis of the Mattis-Bardeen formula for the conductivity of a superconductor. Details on its derivation can be found in \cite{bimonteBCS}. The function $g(\xi)$ can be expressed as:
\be
g(\xi)=\Theta(T_{\rm c}-T) \int_{-\infty}^{\infty} \frac{d \epsilon}{E} \tanh\left(\frac{E}{2 k_B T}\right)\,{\rm Re}[G_+(i \xi,\epsilon)]\;,
\ee
where $\Theta(x)$ is the Heaviside step-function: $\Theta(x)=1$ for $x >0$, and  $\Theta(x)=0$ for $x \le 0$ and
\be
G_+(z,\epsilon)=\frac{\epsilon^2 Q_+(z,E) +(Q_+(z,E)+i \hbar \gamma)A_+(z,E)}{Q_+(z,E) [\epsilon^2-(Q_+(z,E)+i \hbar \gamma)^2]}\;,
\ee
with
\be
E=\sqrt{\epsilon^2+\Delta^2}\;,
\ee
\be
Q_+(z,E)=\sqrt{(E+\hbar z)^2-\Delta^2}\;,
\ee
and
\be
A_+(z,E)=E(E+\hbar z)+\Delta^2\;.
\ee
Here, $\Delta$ is the temperature-dependent gap. From BCS theory \cite{tinkham} one knows that
\be
\Delta=c_1 \,k_B T_{\rm c} \sqrt{1-\frac{T}{T_{\rm c}}}\left(c_2+c_3 \frac{T}{T_{\rm c}}\right)\;.
\ee
where $c_1=1.764$, $c_2=0.9963$ and $c_3=0.7735$

\end{document}